\documentclass[conference]{IEEEtran}
\IEEEoverridecommandlockouts

\usepackage{cite}
\usepackage{amsmath,amssymb,amsfonts}
\usepackage{algorithmic}
\usepackage{graphicx}
\usepackage{textcomp}
\usepackage{xcolor}
\usepackage{booktabs}
\usepackage{hyperref}
\newcommand\liming[1]{}
\def\beps{\boldsymbol{\epsilon}}
\def\ecapa/{ECAPA-TDNN}
\def\ddpovc/{DDPO-VC}
\def\ddpo/{DDPO}
\def\deid/{speaker de-id}
\def\effnet/{EfficientNet}
\def\facodec/{FACodec}
\def\fhsgold/{FHS gold 92}
\def\hifigan/{HiFiGAN}
\def\knnvc/{KNN-VC}
\def\linearvc/{LinearVC}
\def\triaanvc/{TriAAN-VC}
\def\uniaudio/{UniAudio}
\def\valle/{VALL-E}
\def\vevo/{VEVO}
\def\facodec/{FACodec}
\def\wavlm/{WavLM}
\def\whisper/{Whisper}
\def\demenba/{Demenba}
\def\effnet/{EfficientNet}

\usepackage{amsmath,amsfonts,bm}

\def\eqref#1{Eq.~\ref{#1}}

\def\1{\bm{1}}

\def\rvc{{\mathbf{c}}}

\def\rvs{{\mathbf{s}}}

\def\rvx{{\mathbf{x}}}

\def\rvz{{\mathbf{z}}}

\DeclareMathAlphabet{\mathsfit}{\encodingdefault}{\sfdefault}{m}{sl}
\SetMathAlphabet{\mathsfit}{bold}{\encodingdefault}{\sfdefault}{bx}{n}

\def\gN{{\mathcal{N}}}

\newcommand{\E}{\mathbb{E}}

\newcommand{\KL}{D_{\mathrm{KL}}}

\begin{document}

\title{DDPO-VC: Speaker De-Identification via Diffusion Denoising Policy Optimization}

\author{\IEEEauthorblockN{Liming Wang}
\IEEEauthorblockA{\textit{MIT CSAIL}\\
Cambridge, USA\\
limingw@csail.mit.edu}
\and
\IEEEauthorblockN{Cody Karjadi}
\IEEEauthorblockA{
\textit{Boston University}\\
Boston, USA\\
ckarjadi@bu.edu}
\and
\IEEEauthorblockN{Rhoda Au}
\IEEEauthorblockA{\textit{Boston University} \\
Boston, USA\\
rhodaau@bu.edu}
\and
\IEEEauthorblockN{James Glass}
\IEEEauthorblockA{
\textit{MIT CSAIL}\\
Cambridge, USA\\
glass@mit.edu}
}
\maketitle

\begin{abstract}
A key challenge of speaker de-identification is the balance between privacy and utility. Many utility variables, such as the cognitive health status of the speaker, are correlated with the privacy variable, such as the speaker identity, violating the independence assumption held by disentanglement-based approaches. This causes leakage of private information and the loss of useful information for downstream tasks. To tackle this challenge, we propose a general framework, DDPO-VC, for speaker de-identification through reinforcement learning-based post-training with diffusion models. By learning from reward signals and combining knowledge from privacy-focused and utility-focused teachers, our method outperforms various strong \deid/ methods in both privacy preservation and cognitive utility on two commonly used dementia speech benchmarks. 
Please check out our code\footnote{\href{https://github.com/cactuswiththoughts/DDPO-VC}{https://github.com/cactuswiththoughts/DDPO-VC}} and demo \footnote{\href{https://cactuswiththoughts.github.io/SpeakerDeID-Demo/}{https://cactuswiththoughts.github.io/SpeakerDeID-Demo/}}.
\end{abstract}


\section{Introduction}
Speech data is a rich source of information on the Internet and underpins a wide range of applications, from voice assistants and language-learning tools to speech derived metrics in healthcare. However, acoustic properties alone can reveal sensitive speaker information, including personal identity and health status.
Leakage of such private data can expose speakers to identity theft, financial fraud, mental distress, and potential physical harm.

The stakes are particularly high in fields such as healthcare, where leakage of a patient's personally identifiable information (PII) can cause serious harm, including loss of medical resources through medical identity theft and exposure to stigma or discrimination from family members and the public. Older patients and those with cognitive impairment can be especially vulnerable to scams and identity theft.
According to the FTC, more than 1.1 million identity theft cases were reported in the U.S. in 2024 alone, with a significant portion from medical services~\cite{ftc2025consumer_sentinel_2024}.

The goal of a speaker de-identification (\deid/) system is twofold: It should remove speaker-dependent cues from speech while preserving information needed for downstream tasks, such as speech related disease indicator. While de-identification aims to remove or transform identifying information so that data are less directly linked to an individual, it does not necessarily achieve complete anonymization, which requires that the individual can no longer be identified by any reasonably likely means.

A natural way to formulate \deid/ is as a \emph{disentanglement} problem~\cite{Qian2019autovc,wang2021vqmivc}: the model should learn independent representations for 1) speaker identity (privacy) and, 2) non-speaker attributes that support useful downstream tasks such as paralinguistic classification (utility). However, disentanglement has been shown to be generally impossible even with partial labels~\cite{wang2025can}; consequently, existing voice-conversion-based methods cannot guarantee optimal removal of speaker-related information. Moreover, in health-related tasks, latent variables such as disease severity are often correlated with speaker identity, so a more careful balance between privacy and utility need to be struck. This challenge is especially pronounced for dementia speech, where clinically meaningful cues may be entangled with speaker-specific acoustic traits; aggressive removal of identity information can therefore degrade the biomarkers that make the data useful.

In this work, we propose to achieve de-identification via reinforcement learning (RL) post-training of diffusion models (DMs)~\cite{Song2019generative,ho2020denoising}. This approach is motivated by three considerations. First, DMs are powerful generative models capable of producing natural, human-like speech and cloning a speaker's voice from only a few seconds of reference speech~\cite{popov2021gradtts,popov2022diffusionbased,zhang2025vevo}. Compared with discrete-token approaches, DM-based speech generation operates directly on continuous speech representations, avoids quantization-induced information loss, and achieves strong results in text-to-speech synthesis (TTS)~\cite{popov2021gradtts}, voice conversion (VC)~\cite{popov2022diffusionbased,zhang2025vevo}, and \deid/~\cite{huang24_interspeech}. Second, RL provides a generalizable, model-free way to navigate the subtle correlation between privacy and utility variables and optimize the privacy-utility tradeoff tailored to different applications, without explicit modeling assumptions on those variables. By treating privacy and utility objectives as rewards, RL post-training can adapt an existing generative de-identification model to a target application without requiring an explicit factorization of speaker identity and clinical state. Third, while various works have sidestepped the need for RL during post-training of DM via direct preference optimization (DPO)~\cite{wallace2024diffusion}, the latter can only be applied to disentanglement when counterfactual preference pairs are available or can be synthesized easily, making it difficult to disentangle more subtle utility variables such as the cognitive or emotional state of the speaker, especially under low-resource settings.

The main contributions of this paper are threefold: 
\begin{enumerate}
    \item We propose \ddpovc/, a general framework for \deid/ based on RL post-training with DMs that complements existing methods;
    \item We focus on an important and challenging healthcare setting by studying \deid/ of \emph{dementia speech}, a canonical case where the utility variable, dementia, is correlated with the privacy variable, speaker identity, and demonstrate the superiority of our method over prior art.
    \item We provide a careful analysis on the effect of different design choices of our method, and a more diverse set of evaluation metrics for \deid/ tailored to the health domain, which sheds light on new directions for improving existing \deid/ methods.
\end{enumerate}

\section{Related Works}
\liming{Control reference to be less than 2 pages}
Speaker de-id (or speaker anonymization), is closely related to the task of VC~\cite{Qian2019autovc,wang2021vqmivc,park2023triaan,popov2022diffusionbased,wang2023lmvc,ghosh2023emo,ju2024naturalspeech,zhang2025vevo}. Many \deid/ systems have used VC systems as starting points, and perform \deid/ via signal processing-based manipulation~\cite{patino2021mcadams}, VC with multiple target speakers~\cite{fang2019speaker,srivastava2020evaluating,srivastava2020design,turner2020distribution}, noise injection to the speaker embedding~\cite{huang24_interspeech,wang2025async}, vector quantization~\cite{champion2022disentangled,panariello2024nac}, generative speaker modeling~\cite{meyer2022phonetic} and utility variable distillation~\cite{yao2025easy}. Others have prioritized privacy over utility and resource efficiency by using TTS for \deid/~\cite{meyer2022phonetic,meyer2022gan}. Later works have found simple adaptation of VC leads to information leakage, making them susceptible to adversarial attack~\cite{srivastava2020evaluating}, and perform further \deid/ via prosodic information removal~\cite{franzreb25_interspeech,shamsabadi2023dp}. While earlier works have focused on evaluating \deid/ via generic metrics such naturalness, speech recognition word error rate (WER) and signal detection equal error rate (EER), recent approaches have explored more diverse utility metrics such as emotion~\cite{tomashenko2026vp2024results,gaznepoglu2025emotionfail} and pathology~\cite{tayebiarasteh2024pathology} preservation as well as realistic deployment scenario such as defense against adversarial attacks~\cite{srivastava2020evaluating,yao2024svt}.

In parallel, RL with human feedback (RLHF)~\cite{christiano2017deep} and preference alignment-based post-training~\cite{rafailov2023direct} in the context of speech generation is an active research area ~\cite{liu2021reinforcement,hussain2025koeltts,hu2024robust,shankar2024reenact,tian2024preference,zhang2024speechalign,gao2025diffro,yao2025fpo,lin2025alignslm,zhang2025advancing}. 
Beyond plug-and-play applications, recent work attempts to make RL and preference alignment speech-native, sample-efficient, multidimensional, and robust. For example, INTP~\cite{zhang2025advancing} broadens ``hard case'' sampling for TTS intelligibility. FPO~\cite{yao2025fpo} shifts from utterance-level to segment-level supervision.
MPO~\cite{xia2025mpo} regularizes multidimensional alignment. SpeechJudge~\cite{zhang2026speechjudge} and GSRM~\cite{shen2026gsrm} build speech reward models rather than borrowing text judges. FlexiVoice~\cite{chen2026flexivoice}, Re-ENACT~\cite{shankar2024reenact} and VGPO~\cite{liu2026vgpo} show that reward design must track richer control objectives than mere naturalness. While most methods use auto-regressive spoken language models, RL and DPO post-training has also been applied to diffusion~\cite{black2024training,wallace2024diffusion,chen2024dlpo} and flow matching models~\cite{liu2026vgpo}.  

\section{Method}
\begin{figure}
    \centering
    \includegraphics[width=0.5\textwidth]{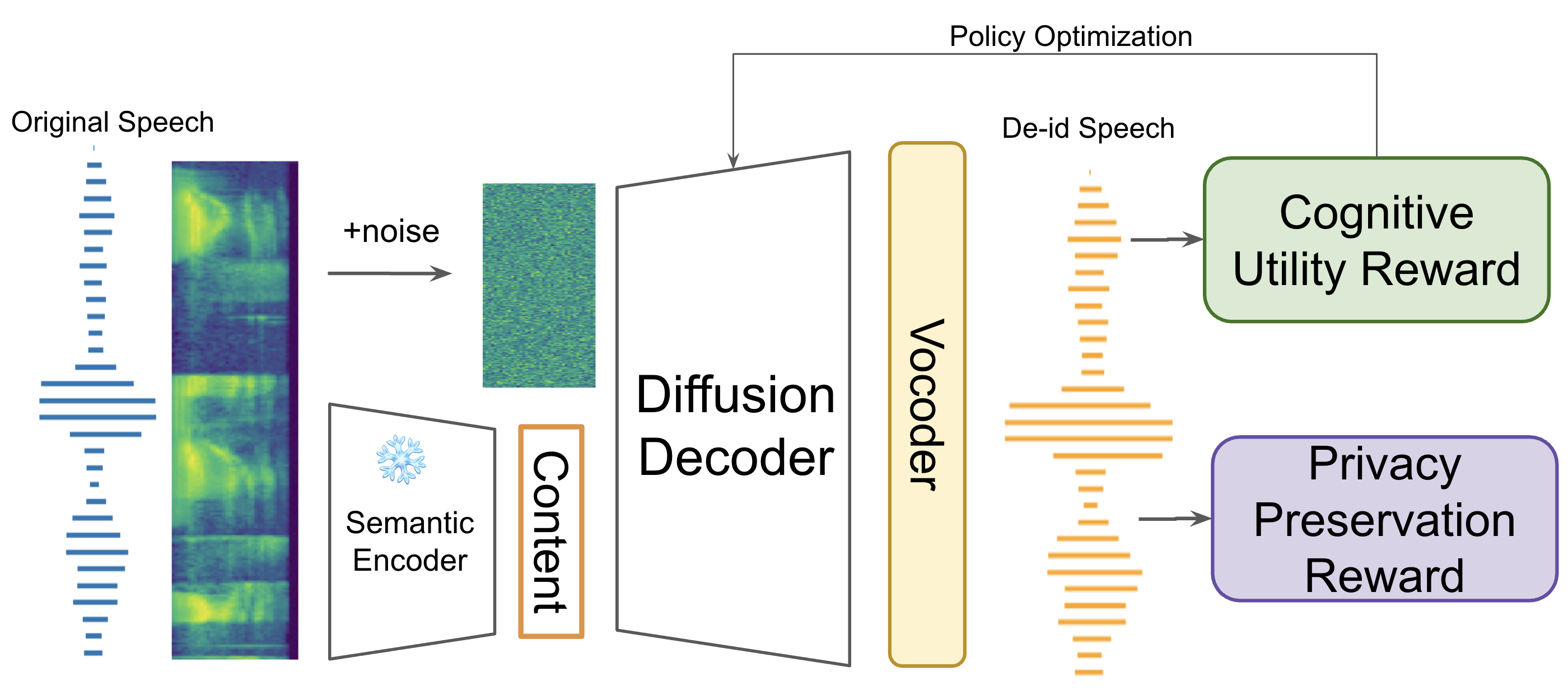}
    \caption{\textbf{Model architecture of \ddpovc/.}\liming{expand}}
    \label{fig:ddpo_vc}
\end{figure}

In this section, we describe \ddpovc/, a \deid/ model that combines conditional diffusion modeling with reinforcement learning (RL) post-training. As shown in Fig.~\ref{fig:ddpo_vc}, our training procedure consists of two stages:  diffusion model pretraining, which learns to synthesize speech from semantic representations, and RL post-training, which explicitly optimizes the privacy-utility tradeoff. 

\subsection{Denoising diffusion probabilistic model}
Given the Mel spectrogram of a spoken utterance $\rvx:=f(\rvc,\rvs)$ with utility variable $\rvc$ and privacy variable, i.e., speaker identity $\rvs$, the goal of \deid/ is to learn a mapping $\rvx' = f(\rvc,\rvs')$ such that the new speaker identity $\rvs'$ is independent of the original identity $\rvs$, while the utility information $\rvc$ is preserved. 

During diffusion model pretraining, we assume access to an \emph{imperfect} semantic encoder $\tilde{\rvc}\approx \tilde{c}(\rvx)$, e.g., a \wavlm/ encoder~\cite{Chen2021WavLM} or other pretrained speech encoders. This encoder is expected to preserve task-relevant information in $\rvc$, but may also retain residual speaker information from $\rvs$. We therefore learn a conditional probability distribution $p(\rvx|\tilde{\rvc})$ from which we can sample a new utterance $\rvx'\sim p(\cdot|\tilde{\rvc})$ that preserves utility while reducing dependence on the original speaker. To this end, we utilize a conditional diffusion model (CDM) based on the denoising diffusion probabilistic model (DDPM)\cite{ho2020denoising}. DDPM gradually corrupts the clean sample $\rvx_0:=\rvx$ into Gaussian noise $\rvx_T$ through a forward diffusion process, and learns the reverse denoising process for generation. Given a variance-preserving noise schedule $\{\alpha_t\}_{t=1}^T$, the forward process can be written as:
\begin{align}\label{eq:forward_proc}
    \rvx_t &:=\sqrt{\bar{\alpha}_t}\rvx_0+\sqrt{1-\bar{\alpha}_t}\beps\sim \gN(\rvx_t; \sqrt{\bar{\alpha}_t}x_0; (1-\bar{\alpha})\mathbf{I}),
\end{align}
where $\bar{\alpha}_t:=\prod_{\tau=1}^t\alpha_{\tau}$ and $\beps_t$ is standard Gaussian. In our implementation, we follow the diffusion schedule and sampler used by the base diffusion VC model~\cite{popov2022diffusionbased}.

For the reverse process, we train a neural denoiser $\epsilon_{\theta}(\rvx_t,\tilde{\rvc},t)$ to predict the injected noise. The pretraining objective, also known as \emph{conditional score matching}~\cite{Song2019generative}, is
\begin{align}\label{eq:score_matching}
    L_{\mathrm{DDPM}}(\theta):=\E_{\rvx_0,t,\beps}\|\epsilon_{\theta}(\rvx_t,\tilde{\rvc},t)-\beps\|^2,
\end{align}
where $t$ is sampled uniformly from $\{1,\ldots,T\}$. After pretraining, de-identified speech is generated by sampling $\rvx'_T\sim\gN(\mathbf{0},\mathbf{I})$ and iteratively applying
\begin{align}
    \rvx'_{t-1} = \frac{1}{\sqrt{\alpha_t}}\left(\rvx'_t-\frac{1-\alpha_t}{\sqrt{1-\bar{\alpha}_t}}\epsilon_{\theta}(\rvx'_t,\tilde{\rvc},t)\right)+\sigma_t^2 \rvz,
\end{align}
where $\rvz\sim\gN(\mathbf{0},\mathbf{I})$ and $\sigma_t>0$ controls the reverse-process variance.

\subsection{Denoising diffusion policy optimization}
Although the pretrained CDM can synthesize speech from semantic representations, residual speaker information in $\tilde{\rvc}$ may still leak into the generated sample. We therefore post-train the base diffusion model with RL. In this stage, a complete reverse denoising trajectory is treated as a rollout, and the final generated utterance is scored by black-box reward models. When these reward models operate on waveforms, the generated Mel-spectrogram is first converted to waveform by the vocoder; we keep the notation $\rvx'$ for simplicity.

To encourage privacy, a \emph{privacy teacher} is used to reward the generated speech for moving away from the original speaker. Specifically, a pretrained speaker verifier $\tilde{s}(\cdot)$, e.g., ECAPA-TDNN\cite{desplanques2020ecapa}, maps the original and generated utterances to speaker embeddings. The speaker reward is the cosine distance between these embeddings:
\begin{align}
    r_{\mathrm{speaker}}(\rvx'):=1-\cos(\tilde{s}(\rvx), \tilde{s}(\rvx')).
\end{align}
A larger value indicates lower similarity to the source speaker. To preserve cognitive utility, we use a pretrained dementia classifier as a \emph{cognitive utility teacher}. Let $\hat{p}(y|\rvx')$ denote its posterior probability for dementia label $y$. Given the true label $y^*$ of the source utterance, the dementia reward is
\begin{align}
    r_{\mathrm{dementia}}(\rvx'):=\hat{p}(y^*|\rvx').
\end{align}
The total reward is a weighted sum of utility and privacy rewards:
\begin{align}
    r(\rvx'):=r_{\mathrm{dementia}}(\rvx')+\lambda_{\mathrm{speaker}} r_{\mathrm{speaker}}(\rvx'),
\end{align}
where $\lambda_{\mathrm{speaker}} > 0$ controls the relative strength of the privacy--utility tradeoff.

To finetune the base CDM with parameters $\theta_{\mathrm{base}}$, we employ denoise diffusion policy optimization (DDPO)\cite{black2024training}. DDPO maximizes a KL-regularized objective that rewards generated samples while constraining the post-trained policy to remain close to the base model:
\begin{align}\label{eq:entropy_regularized_reward}
    \E_{\rvx'\sim p_{\theta}(\cdot|\tilde{\rvc})}[r(\rvx')]
    -\beta\KL\!\left(p_{\theta}(\cdot|\tilde{\rvc})\,\|\,
    p_{\theta_{\mathrm{base}}}(\cdot|\tilde{\rvc})\right),
\end{align}
where $\beta > 0$ mitigates reward hacking and distributional drift. Following \cite{black2024training}, we optimize Eq.~\ref{eq:entropy_regularized_reward} with a denoising-score-matching surrogate:
\begin{multline}\label{eq:ddpo}
    L_{\mathrm{DDPO}}(\theta)=
    \E_{\rvx',t,\beps}\left[r(\rvx')
    \|\epsilon_{\theta}(\rvx_t,\tilde{\rvc},t)-\beps\|^2\right]+\\
    \beta\E_{\rvx',t,\beps}
    \|\epsilon_{\theta}(\rvx_t,\tilde{\rvc},t)-
    \epsilon_{\theta_{\mathrm{base}}}(\rvx_t,\tilde{\rvc},t)\|^2,
\end{multline}
where $\rvx'$ is a generated rollout sample. The first term increases the likelihood of high-reward denoising trajectories, and the second term regularizes the post-trained denoiser toward the frozen base model.

In practice, raw rewards can have different scales and may destabilize RL training. We therefore replace $r$ in Eq.~\ref{eq:ddpo} with a normalized and clipped reward. For the $k$-th rollout, we compute
\begin{align*}
    \tilde{r}(\rvx^{(k)}) = \mathrm{clip}\!\left(
    \frac{r(\rvx^{(k)})-\mu_{k-1}}{\nu_{k-1}},
    -\delta,+\delta\right),
\end{align*}
where $\mu_{k-1}$ and $\nu_{k-1}$ are exponential moving averages of the mean and standard deviation of previous rewards, computed with decay rate $0.98$. In addition to fixed teachers, we also consider a trainable utility teacher that is updated alternately with the CDM, yielding an actor-critic-style training procedure~\cite{liu2026vgpo}.

\section{Experiment}
\subsection{Dataset}
We pretrain \ddpovc/ on an 800-hour subset of the FHS dataset~\cite{wang2025recognizing} dementia speech dataset, balanced between healthy and dementia speech, and follow the data preprocessing procedure in \cite{wang2025recognizing} and keep only the patient speech. 
We also follow the training hyperparameters and diffusion schedules and sampler in \cite{popov2022diffusionbased}.

We post-train and evaluate \ddpovc/ on two dementia classification datasets. The first is ADReSS~\cite{Luz2020_alzheimer_adress}, a dementia speech corpus with 78 healthy and 78 participants with Alzheimer's Disease. Each participant has an average of 25 speech segments, and each segment is shorter than 10 seconds. The standard split is used. The second is \fhsgold/\cite{al-hanai-etal-2018-role}, a larger but noisier dataset containing 72 healthy participants and 20 participants with cognitive impairment, each recorded during a one-hour neuro-psychological test with an examiner. To avoid interference from the examiner's speech, we used the diarization timestamps provided in the dataset to filter out the examiner's speech. We also randomly split the remaining participant speech to make the training and test label distribution approximately the same.

\subsection{Evaluation metrics}
We evaluate \deid/ performance using equal error rate (EER), computed with a pretrained ECAPA-TDNN~\cite{desplanques2020ecapa}. Higher EER indicates stronger speaker de-identification. We evaluate cognitive utility using area under the ROC curve (AUC) for dementia classification on the de-identified speech. For this evaluation, we implement a classifier similar to \cite{Li2023-alzheimer-whisper}, using temporal attention pooling and weighted sum over all the layer embeddings from a \whisper/-small encoder~\cite{Radford2023-whisper}. 

We report two utility settings. In the zero-shot setting, the dementia classifier is trained only on original speech and evaluated directly on \deid/ speech. In the finetuned (ft) setting, the dementia classifier is additionally adapted to \deid/ speech before evaluation. These two settings capture different deployment conditions, depending on whether downstream models can be retrained on de-identified data. For ADReSS, the \whisper/ classifier performed poorly on \deid/ speech in the zero-shot setting, so we use an \effnet/-based approach for that setting~\cite{tan2019efficientnet,dawalatabad-etal-2022-detecting}. Because \whisper/ processes at most 30 seconds of audio at once, we aggregate predictions over 30-second segments by averaging segment-level probabilities. Finally, we measured perceptual quality and naturalness using UTMOS, an automated version of the mean opinion score~\cite{baba2024utmosv2}.

\subsection{Experimental setup}
We used a diffusion-based VC system similar to \cite{huang24_interspeech} as the base model. The semantic encoder consists of the first 18-th layers of a frozen \wavlm/ model, and the diffusion model has 1.19M trainable parameters. We remove speaker conditioning to focus on \deid/. All source audio is resampled to 16kHz and converted to 80-dimensional Mel-spectrogram with a window size of 1280, and a skip size of 320 is used for DDPM training to match the frame rate of the \wavlm/ encoder. A \hifigan/~\cite{kong2020hifi} vocoder trained to match the same Mel-spectrogram configuration converts generated spectrogram back to waveforms. 

For pretraining and post-training, we trained the model on 10-second audio segments with batch sizes of 32 and 16, respectively. We pretrained the base model for 100 epochs and post-trained the model for 1000 DDPO steps. For the cognitive utility teacher, we evaluate both the \whisper/-based and the \effnet/-based dementia classifiers described above. For the privacy teacher, we used the \ecapa/ speaker verifier. During DDPO, we use 50 diffusion steps to generate online samples, set the trust-region clipping threshold to $\delta=0.5$, and use KL regularization weight $\beta=0.2$. All models are trained on two NVIDIA A6000 GPUs.

We compared our approach with a variety of strong, publicly available baselines, including VC models based on nearest-neighbor matching (KNN-VC~\cite{baas2023knnvc}, Linear-VC~\cite{kamper2025}), adaptive attention (\triaanvc/~\cite{park2023triaan}), neural codec (\facodec/~\cite{ju2024naturalspeech} and \vevo/~\cite{zhang2025vevo}), and a text-to-speech synthesis (TTS)-based method (\valle/\footnote{https://github.com/Plachtaa/VALL-E-X}~\cite{wang2023neural}). For TTS baselines, we synthesize speech from the ground-truth transcript with a target speaker different from the source speaker randomly selected from the dataset. We also compared and combined \ddpovc/ with the alternative post-training techniques for CDM based on diffusion direct preference optimization (DPO)~\cite{wallace2024diffusion}, using real samples with different dementia labels and synthetic samples from the base model ranked by the speaker teacher as preference pairs.

\begin{figure}
    \centering
    \includegraphics[width=0.5\textwidth]{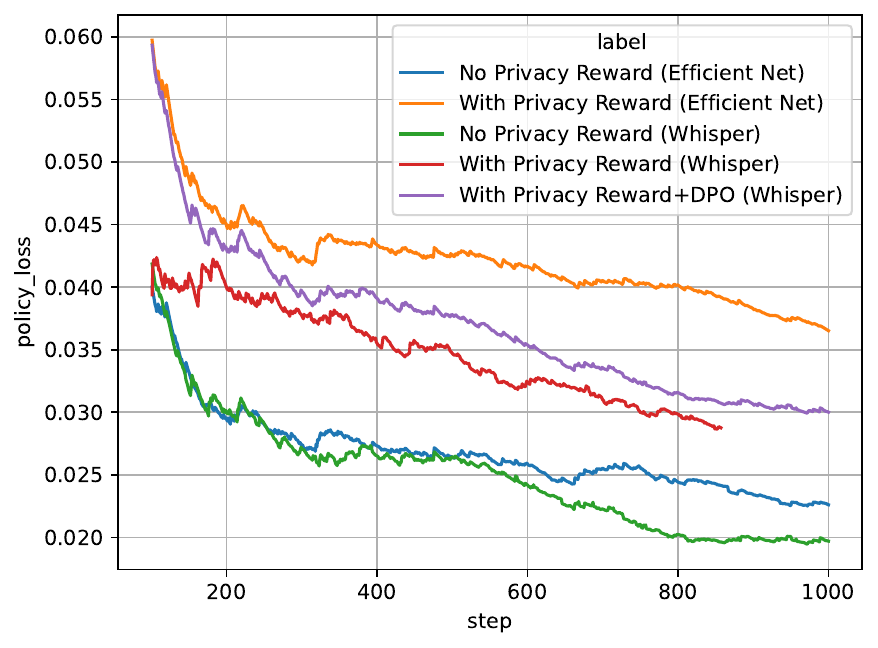}
    \caption{\textbf{Training loss convergence for \ddpovc/ variants on ADReSS.} The curves compare RL objectives and teacher configurations used during post-training.}
    \label{fig:training_curve}
\end{figure}

\begin{table*}[t]
\centering
\caption{\textbf{AUC, EER, and UTMOS results for different \deid/ methods on ADReSS}. Boldface and underlining denotes the best and the second-best results respectively.}
\begin{tabular}{lcccc}
\toprule
 & AUC (zs, $\uparrow$) & AUC (ft, $\uparrow$) & EER ($\uparrow$) & UTMOS ($\uparrow$)\\
\midrule
original & 0.85 & 0.85 & 0.13 & 1.99\\
\knnvc/~\cite{baas2023knnvc} & 0.66 & 0.85 & 0.37 & \underline{2.22}\\
\triaanvc/~\cite{park2023triaan} & 0.75 & 0.58 & 0.38 & 2.13\\
\valle/~\cite{wang2023neural} & 0.75 & 0.78 & \textbf{0.46} & \textbf{2.63}\\
\linearvc/~\cite{kamper2025} & 0.70 & \textbf{0.89} & 0.28 & 2.00\\
\vevo/~\cite{zhang2025vevo} & 0.67 & 0.85 & 0.40 & 1.95\\
\facodec/~\cite{ju2024naturalspeech}  & 0.66 & 0.74 & 0.32 & 1.32 \\
\midrule
\ddpovc/ (base)  & 0.57 & 0.75 & 0.42 & 1.73 \\
\ddpovc/ (fixed reward) & \underline{0.76} & 0.78 & 0.42 & 1.74\\
\ddpovc/ (trainable reward) & \textbf{0.77} & \underline{0.87} & \underline{0.43} & 1.98\\
\bottomrule
\end{tabular}
\label{tab:deid_results_adress}
\end{table*}

\begin{table*}[t]
\centering
\caption{AUC, fine-tuned AUC, and EER results for different \deid/ methods on \fhsgold/.\liming{TODO parameter size}}
\begin{tabular}{lcccc}
\toprule
 & AUC (zs, $\uparrow$) & AUC (ft, $\uparrow$) & EER ($\uparrow$) & UTMOS ($\uparrow$)\\
\midrule
original & 0.79 & 0.79 & 0.34 & 1.32\\
\knnvc/~\cite{baas2023knnvc} & 0.55 & 0.55 & \underline{0.47} & 1.30\\
\linearvc/~\cite{kamper2025} & 0.61 & 0.81 & 0.41 & 1.51\\
\triaanvc/~\cite{park2023triaan} & \textbf{0.82} & 0.85 & 0.35 & \underline{1.59}\\
\valle/~\cite{wang2023neural} & \textbf{0.82} & 0.85 & 0.35 & \textbf{1.73}\\
\vevo/~\cite{zhang2025vevo} & 0.56 & \underline{0.87} & 0.46 & 1.37\\
\facodec/~\cite{ju2024naturalspeech} & 0.62 & \textbf{0.92} & 0.44 & 1.30\\
\midrule
\ddpovc/ (base) & 0.65 & 0.69 & 0.41 & 1.32\\
\ddpovc/ (fixed reward) & \underline{0.66} & 0.83 & 0.43 & 1.33\\
\ddpovc/ (trainable reward) & 0.56 & \textbf{0.92} & \textbf{0.50} & 1.42\\
\bottomrule
\end{tabular}
\label{tab:deid_results_fhs}
\end{table*}
\subsection{Results}
Fig.~\ref{fig:training_curve} shows the training loss curves for different reward models on ADReSS. \ddpovc/ with the \effnet/ dementia teacher is less stable than with the \whisper/ teacher, suggesting that the richer linguistic representations in \whisper/ provide a smoother utility signal. Adding DPO also slows convergence, consistent with the fact that preference pairs are easier to define for speaker dissimilarity than for subtle cognitive preservation.

Table~\ref{tab:deid_results_adress} reports the overall results on ADReSS. \ddpovc/ achieves the best zero-shot AUC among all methods, reaching 0.76 with a fixed reward and 0.77 with a trainable reward. However, \linearvc/ has substantially weaker privacy preservation, with an EER of 0.28 compared with 0.43 for \ddpovc/. This suggests that linear latent-space conversion may not adequately separate privacy- and utility-related information, whereas RL post-training can directly optimize their tradeoff.
Among the baselines, \valle/ achieves the highest EER of 0.46, but it relies on ground-truth text and can discard prosodic and paralinguistic cues that are important for health-related utility. In contrast, \ddpovc/ uses only untranscribed speech and achieves comparable privacy with better dementia utility, making it more suitable for low-resource settings. We also observe that \ddpovc/ with a trainable reward nearly matches the UTMOS of the original speech. Some VC and TTS baselines achieve even higher UTMOS, likely because they denoise the source audio or synthesize cleaner speech from text. However, their higher naturalness does not consistently translate into higher cognitive utility, indicating that perceptual quality alone is insufficient for evaluating health-oriented \deid/.
RL post-training substantially improves \ddpovc/ over the base model. On ADReSS, fixed-reward post-training improves zero-shot AUC by 10\% relative and finetuned AUC by 4\% relative, while preserving EER. Using a trainable reward yields an additional 12\% relative gain in finetuned AUC and a 14\% relative gain in UTMOS over the base model. We hypothesize that the trainable reward remains better calibrated as generated speech moves away from the original data distribution, whereas a fixed reward model may overfit to the original speech distribution.

Table~\ref{tab:deid_results_fhs} shows the results on \fhsgold/. In the zero-shot setting, \ddpovc/ outperforms four of the six baselines and is behind only \triaanvc/ and \valle/, both of which have lower EER and thus weaker speaker privacy. In the finetuned setting, \ddpovc/ with a trainable reward reaches the best AUC, matching \facodec/ at 0.92 while improving EER by 14\% relative. As on ADReSS, fixed-reward RL post-training improves both finetuned AUC and EER over the base model, and the trainable reward provides an additional 11\% relative gain in finetuned AUC. Unlike on ADReSS, the trainable reward reduces zero-shot AUC on \fhsgold/. This suggests that RL post-training shifts the generated speech distribution more strongly on the noisier \fhsgold/ corpus, making evaluation harder for a dementia classifier trained only on original speech. Nevertheless, our method achieves similar or better UTMOS than the original speech, likely due to the  denoising effect of diffusion generation. \facodec/ also performs surprisingly well on \fhsgold/, unlike on ADReSS; one possible explanation is that discrete codec representations are more robust under the domain shift and noise level present in \fhsgold/.

\subsection{Ablations}
We study three design choices in \ddpo/ training: the cognitive utility teacher, the post-training objective, and the relative weight of the privacy reward.

\begin{table}[ht]
    \centering
    \caption{The effect of teacher type on \deid/ performance on ADReSS for \ddpovc/ (fixed reward).}
    \begin{tabular}{lcccc}
    \toprule
    Teacher type & AUC (zs,$\uparrow$) & AUC (ft,$\uparrow$) & EER ($\uparrow$) & UTMOS ($\uparrow$)\\
    \midrule
    \whisper/ & \textbf{0.76} & \textbf{0.78} & \textbf{0.42} & \textbf{1.74} \\
    \effnet/ & 0.72 & 0.75 & 0.42 & 1.69\\
    \bottomrule
    \end{tabular}
    \label{tab:effect_of_dementia_teacher}
\end{table}

Table~\ref{tab:effect_of_dementia_teacher} compares cognitive utility teachers. Compared with \effnet/, the \whisper/ teacher performs similarly or better across all metrics, likely because it captures linguistic cues that are informative for cognitive status.

\begin{table}[ht]
    \centering
    \caption{The effect of different post-training techniques on \deid/ performance on ADReSS.}
    \begin{tabular}{lcccc}
    \toprule
    Post-training & AUC (zs,$\uparrow$) & AUC (ft,$\uparrow$) & EER ($\uparrow$) & UTMOS ($\uparrow$)\\
    \midrule
    \ddpo/ & \textbf{0.76} & \textbf{0.78} & 0.42 & \textbf{1.74}\\
    DPO & 0.57 & 0.60 & \textbf{0.49} & 1.71\\
    \ddpo/+DPO & 0.68 & 0.76 & 0.41 & 1.72\\
    \bottomrule
    \end{tabular}
    \label{tab:effect_of_dpo}
\end{table}

Table~\ref{tab:effect_of_dpo} compares post-training objectives. Diffusion DPO achieves the highest EER but performs significantly worse than \ddpo/ on both AUC metrics, suggesting that DPO is effective for increasing speaker dissimilarity but less effective for preserving dementia-related information. This is expected because preference pairs with different speaker identities are easier to construct than counterfactual pairs that differ only in cognitive status. Combining DPO with \ddpo/ using equal loss weighting does not improve performance, so better integration of preference optimization and reward-based post-training remains future work.

\begin{table}[ht]
    \centering
    \caption{Effect of Reward Type on AUC, fine-tuned AUC, and EER results on ADReSS.}
    \begin{tabular}{lcccc}
    \toprule
    $\lambda_{\mathrm{speaker}}$ & AUC (zs,$\uparrow$) & AUC (ft,$\uparrow$) & EER ($\uparrow$) & UTMOS ($\uparrow$)\\
    \midrule
    0  & \textbf{0.77} & \textbf{0.87} & \textbf{0.43} & \textbf{1.98}\\
    0.05 & 0.66 & 0.84 & 0.37 & 1.80\\
    0.5 & 0.66 & 0.78 & 0.37 & 1.80\\
    \bottomrule
    \end{tabular}
    \label{tab:eff_reward_type}
\end{table}

Finally, Table~\ref{tab:eff_reward_type} studied the privacy reward weight. Increasing $\lambda_{\mathrm{speaker}}$ degrades performance across metrics. This suggests that the semantic encoder may already remove substantial speaker information, or that the frozen privacy teacher is susceptible to reward hacking under stronger weighting. We leave a deeper analysis of this behavior to future work.

\section{Conclusion}

We introduced \ddpovc/, an RL-post-trained diffusion framework for \deid/ tailored to dementia speech. Across two realistic dementia datasets, \ddpovc/ improves the privacy-utility tradeoff, achieving strong cognitive-utility and speaker-privacy results relative to competitive VC, codec, and TTS baselines. Our analysis also shows that conventional metrics such as naturalness and speaker similarity alone are insufficient: health-oriented \deid/ should be evaluated using task-specific utility, privacy, intelligibility, naturalness, and distributional preservation in tandem. Because the framework only requires reward signals, it can be extended to other audio domains, utility variables, and generative architectures. Future work will evaluate robustness against adaptive re-identification attacks, develop stronger privacy reward models, and design training strategies that mitigate reward hacking.

\bibliographystyle{IEEEtran}
\bibliography{slt2026.bib,refs_deid,refs_rl_speech}

\end{document}